\documentclass[iop]{emulateapj}
\usepackage{graphicx}
\usepackage{epstopdf}
\usepackage{amssymb,amsmath}

\slugcomment{}
\shorttitle{Shocks in Tilted Accretion Flows}
\shortauthors{Generozov et al.}

\begin{document}

\title{Physical Properties of the Inner Shocks in Hot, Tilted Black Hole
Accretion Flows}

\author{Aleksey Generozov}
\affil{Department of Astronomy, Columbia University, 550 West 120th Street, New York, NY 10027}

\author{Omer Blaes}
\affil{Department of Physics, University of California, Santa Barbara,
   Santa Barbara CA 93106}

\author{P. Chris Fragile\footnote{KITP Visiting Scholar, Kavli Institute for Theoretical Physics, Santa Barbara, CA.}}
\affil{Department of Physics and Astronomy, College of Charleston,
   Charleston, SC 29424}

\and

\author{Ken B. Henisey}
\affil{Natural Science Division, Pepperdine University, Malibu, CA 90263}

\begin{abstract}
Simulations of hot, pressure supported, tilted black hole accretion flows,
in which the angular momentum of the flow is misaligned with the black hole
spin axis, can exhibit two non-axisymmetric shock structures in the inner
regions of the flow.  We analyze the strength and significance of these shock
structures in simulations with tilt angles of 10 and 15 degrees.
By integrating fluid trajectories in the simulations through the shocks,
and tracking the variations of fluid quantities along these trajectories,
we show that these shocks are strong, with substantial compression ratios,
in contrast to earlier claims.  However, they are only moderately relativistic.
 We also show that the two density enhancements resembling flow streams
in their shape are in fact merely post-shock compressions, as fluid
trajectories cut across, rather than flow along, them.  The dissipation associated
with the shocks is a substantial fraction ($\simeq3-12$ percent) of the rest
mass energy advected into the hole, and therefore comparable to the
dissipation expected from turbulence.  The shocks should therefore make
order unity changes in the observed properties of black hole accretion
flows that are tilted.
\end{abstract}


\section{Introduction}
\label{sec:introduction}

Much has been learned concerning the dynamics of black hole accretion flows
with magnetorotational (MRI, \citealt{bal91,bal98}) turbulence in recent
years through the use of global, general relativistic magnetohydrodynamical
(GRMHD) simulations.  In cases where the effects of black hole spin are
included, most of these simulations have assumed that the angular momentum of
the accretion flow is aligned with the black hole spin axis (e.g.
\citealt{dev03, gam03, kro05, nob09, pen10}).  This has been done largely
for simplicity, and because turbulent stresses interacting with
Lense-Thirring precession have long been believed to
align the inner accretion disk with the black hole spin
axis \citep{bar75}, at least when the disk is geometrically thin
and turbulent stresses are approximately isotropic.
How this alignment might really work in geometrically thin disks with
MRI turbulence has only begun to be studied very recently \citep{sor13b}.

For geometrically thick disks, tilts and warps are generally thought to
propagate as waves rather than diffuse in the disk \citep{pap95}.
It is then far from clear
that an alignment of the inner disk would take place.  In fact, GRMHD
simulations of misaligned or tilted thick flows extending out to a finite
radius undergo global rigid body precession at a frequency determined by
the ratio of the radially integrated Lense-Thirring
torque to the total angular momentum of the flow \citep{fra07}, but
independent of the actual tilt angle.  On the other hand, if magnetic fields
in the vicinity of the black hole can build up to exert forces that are
sufficient to resist accretion, then rapid alignment of the flow by these
magnetic fields can take place \citep{mck13}.

Given that the angular momentum of the fuel source for the accretion disk
around a black hole is unlikely to have any causal connection to the black
hole spin, it would appear that misaligned or tilted accretion flows
could be quite common in nature.  This
is especially true in the Galactic center source Sgr~A$^\star$, which
is fueled by winds from surrounding stars (e.g. \citealt{cua06}) and
orbiting interstellar gas clouds \citep{gil12}, neither of which
can possibly know about the spin axis of the hole.  It would be an
amazing coincidence if this particular radiatively inefficent flow would
happen to be aligned with the black hole spin.  Misalignments are also likely
to be common for accreting supermassive black holes as new sources of
fuel enter the black hole's gravitational sphere of influence.  The
consequences of this for the evolution of black hole spin in active galactic
nuclei has received much recent attention (e.g. \citealt{kin06,vol07}).

In principle, a long-lived fuel source with stable angular momentum, e.g.
in an X-ray binary,
could torque the black hole spin into alignment given sufficient time,
but estimates for how long this would take (e.g. \citealt{mar08})
depend on the uncertain physics
of Lense-Thirring precession combined with MRI turbulence in a geometrically
thin disk.  Recent hydrodynamic and MHD simulations of geometrically
thin, warped disks \citep{sor13a,sor13b} have begun to shed light on this
physics, emphasizing the importance of transonic flows that mix angular
momenta when warps are nonlinear, and the importance of the fact that
magnetorotational turbulent stresses are anisotropic, in contrast to what
is generally assumed in alpha-based modeling of warps.

The observational evidence for alignments or misalignments in black hole
X-ray binaries is mixed.  Measurements of the orientation of the radio jets
\citep{hje95} and the binary orbital angular momentum \citep{gre01} of
the microquasar GRO~J1655-40 suggest a misalignment of greater than 15
degrees.  On the other hand, modeling of the observed jet kinematics
in XTE~J1550-564 by \citet{ste12} gives a jet orientation that may be
consistent with the binary orbital angular momentum.  Given that only the
inclination of the binary angular momentum with respect to the line of
sight has been measured, not the position angle, they conclude
on a statistical basis that the misalignment with the jet axis is less than 
$12^\circ$ (90\% confidence).  As we shall see here, though, even
a tilt as small as this can have interesting consequences.

Tilted black hole accretion flows that are hot and geometrically thick
exhibit significant dynamical differences compared to their untilted
counterparts.  Neglecting the small stochastic
velocities associated with turbulence, the orbits of fluid elements in
tilted flows are eccentric, not circular.  At least for modest tilts, the
major axes of these eccentric orbits are parallel to the line of nodes
between the disk midplane and black hole equatorial plane, and the
orientation of these orbits flips by $180^\circ$ across the disk
midplane \citep{fra08}.  The eccentricity of the orbits increases
as one moves inward in radius, implying a convergence of fluid trajectories
near their apocenters \citep{iva97}.  As a result, a pair of standing
shocks form in the flow on opposite sides of the black hole along the
line of nodes, but at high latitude \citep{fra08}.
The innermost region of the flow near the black hole is also significantly
nonaxisymmetric, with two high density arms connecting high latitude
regions of the larger thick disk to the black hole \citep{fra07}.  Various
measures of the average inner truncation radius of the
disk are approximately independent of black hole spin, in
contrast to the case of untilted accretion disks where these measures of
truncation radius decrease with increasing black hole spin \citep{fra09,dex11}.
High frequency variability that is enhanced compared to untilted flows is
expected because of the interaction of transient clumps and spiral acoustic
waves with the nonaxisymmetric standing shocks \citep{hen09,hen12}.

A tilted accretion flow necessarily breaks axisymmetry, and so it is not
surprising that nonaxisymmetric structures exist.
However, the origin and nature of the high density arms in the
innermost region is not entirely clear.  \citet{fra07} asserted that
these arms were ``plunging streams", based in part on the fact that
they appeared to originate from high latitude points that coincided
with the generalized innermost stable circular orbit (ISCO) surface
for inclined geodesics orbiting at constant coordinate radius.
The Mach numbers
of the shocks were also claimed to be very modest and described as weak
\citep{fra08}, though actual measurements of the Mach numbers over the
complete shock surfaces were not carried out.  The shocks clearly
exerted significant torques on the flow and of course generated entropy
\citep{fra08,dex11,hen12}, but a calculation of how much dissipation of
accretion power was actually associated with them was not carried out.

It is the purpose of this paper to present a more complete analysis of
the dynamics of the high density arms in the innermost regions, and both
the dynamics and dissipation of the shocks.  We find that the density
arms are not in fact coherent streams of material, but rather post-shock
compressed regions in the flow (see also \citealt{hen12}).
Moreover, the shocks themselves are
strong, not weak, with upstream Mach numbers as high as $\simeq4.7-6$,
depending on time, for a $15^\circ$
tilt.  They also dissipate approximately 3 to 12 percent of the available
accretion power.  The shocks in a tilted accretion flow therefore produce very
substantial changes in both the dynamics and thermodynamics compared
to an untilted accretion flow, which will likely result in
correspondingly significant changes in their observed properties.

This paper is organized as follows.  In section~\ref{sec:simulations} we
provide a brief overview of the simulations used in our analysis.
We measure the fraction of accretion power dissipated in the shocks in
section~\ref{sec:shockdissipation}, and then explore the dynamical connection
between the shocks and the high density arms in the innermost regions
in section~\ref{sec:shockdynamics}.  We present the variation of shock
strength with spatial position and with tilt angle
in section~\ref{sec:shockstrength}.  We summarize our
conclusions and briefly discuss the implications in
section~\ref{sec:conclusions}.  Throughout this paper, we use a metric
with signature $(-+++)$, and except where otherwise specified,
we use units where the speed of light and Newton's gravitational constant
are unity:  $c=G=1$.

\section{Simulations}
\label{sec:simulations}

The GRMHD simulations used in our study are 0910h and 0915h,
described and reported on previously by \citet{fra07} and \citet{hen12},
and generated using the {\it Cosmos}++ code \citep{ann05}.
Both assumed a Kerr spacetime with dimensionless black hole spin $a/M=0.9$,
and used Kerr-Schild coordinates \citep{fon98}, rotated by
the desired tilt of the accretion flow \citep{fra05,fra07a}.  The coordinate
radius of the horizon for this black hole spin is $1.44M$.

The simulations were initialized with a torus with pressure maximum at
coordinate radius $r=25M$.  A circular geodesic in the equatorial plane
of the black hole at this coordinate radius has an orbital period
$t_{\rm orb}\simeq791M$, and this is close to the actual orbital
period of the fluid at the initial pressure maximum of the tilted torus.
The torus was initialized with poloidal magnetic field lines
along isobaric surfaces, together with random perturbations which seeded
the growth of the MRI.

The flow was evolved using the internal energy scheme of {\it Cosmos}++
with an artificial viscosity that we review in more detail in
section~\ref{sec:shockdissipation} below.  Total energy is not conserved
in these simulations, as magnetic and kinetic energy losses occur at the
grid scale and are not recaptured as in simulations that use a total
energy scheme.

The spatial grid used in the simulations consisted of a Kerr-Schild polar
coordinate mesh with $32\times32\times32$ zones at the base level encompassing
all $4\pi$ steradians.  In addition there were two layers of refinement with
twice the resolution in each refinement, resulting in a peak resolution
equivalent to a simulation with $128\times128\times128$ zones.
This peak resolution was achieved over the region $1.41M\le r\le120M$,
$0.111\pi\le\vartheta\le0.889\pi$ and $0\le\varphi<2\pi$ in the tilted
Kerr-Schild coordinates.\footnote{The $0.075\pi\le\vartheta\le0.925\pi$
limits stated in \citet{fra07} are incorrect, and actually refer to the limits
of the concentrated latitude coordinate $x_2$ in that paper.} In other
words, the peak resolution covered most of the simulation domain except
the regions near the polar axes of the tilted Kerr-Schild coordinate system.
In addition, grid zones were spaced logarithmically in the radial direction,
nonuniformly in the polar direction with a concentration toward the midplane
of the original tilted torus, and uniformly in the azimuthal direction.
Much more detail on this gridding and other technical aspects of these
simulations can be found in \citet{fra07}.

We focus here on data in the time range from 4~$t_{\rm orb}$,
after which MRI turbulence is well-established in the inner regions of the
flow near the black hole, to 10~$t_{\rm orb}$, the simulation duration.
All fluid variable data from the simulations was written out to files 200
times per orbit at the pressure maximum radius $25M$ of the initial torus.
This corresponds to a time of $\simeq3.955M$ between data dumps.

\section{Dissipation Associated with the Shocks}
\label{sec:shockdissipation}

As we noted above, the {\it Cosmos}++ simulations we are analyzing were
performed
by integrating the internal energy equation, and not a total energy equation.
As a result, energy is not fully conserved because of losses of magnetic and
kinetic energy at the grid scale.  Simulations such as these therefore mimic an
accretion flow wherein accretion power is lost by cooling, but in a completely
uncontrolled and artificial way.  Similar simulations have been done to explore
the dynamics of MRI turbulence in untilted accretion flows
(e.g. \citealt{dev03}), but using them
to explore the dissipation of this turbulence is clearly problematic.  One
approach has been to adopt proxies such as the square of the electric current
density which might track resistive dissipation \citep{bec06}, or the
electromagnetic stress tensor combined with gradients in fluid velocity
\citep{bec08}.

The one aspect of the dissipation that is explicitly treated in our simulations,
however, is the dissipation associated with the shocks.  The artificial
viscosity used in {\it Cosmos}++ is designed to both numerically resolve
these shocks and to capture the mechanical energy dissipated into gas
internal energy.  We can therefore use this artificial viscosity dissipation
to measure the likely contribution to the overall dissipation of accretion
power by the large scale shocks formed in tilted accretion flows.  We note,
however, that because fluid pressure is fundamentally related to dissipative
heating in any real accretion flow, energy conserving simulations with or
without cooling prescriptions may produce quantitative differences in shock
strength and dissipation from the ones we present here.  For example, in
an energy conserving, radiatively inefficient flow with no significant
cooling, the flow will be hotter than that simulated here, and the Mach
numbers of the fluid motions may therefore be smaller, possibly resulting
in somewhat weaker shocks.

The internal energy equation used in our GRMHD simulations is
\begin{equation}
\partial_tE+\partial_i(EV^i)=-P\partial_tW-(P+Q)\partial_i(WV^i),
\label{eq:internalenergy}
\end{equation}
where $V^i\equiv u^i/u^t$ is the coordinate three-velocity, $u^\alpha$ is the
fluid four velocity, $W=(-g)^{1/2}u^t$ is the relativistic boost factor,
$Q$ is the artificial viscosity, $P$ is the fluid rest frame gas pressure,
$E$ is related to the fluid rest frame internal energy density
$e=P/(\Gamma-1)$ (excluding rest mass) by $E=We$, and $\Gamma$ is the adiabatic
index of the gas (assumed to be 5/3 in our simulations).  The left hand side of this equation can be rewritten as
\begin{equation}
\partial_tE+\partial_i(EV^i)=(-g)^{1/2} \nabla_\alpha (u^{\alpha} e)
\end{equation}
Consequently, when integrated over the duration and volume of the
simulation, this term represents the total change in the internal energy
of the fluid.  The integral of the right hand side of the internal energy
equation (\ref{eq:internalenergy}) consists of a pressure term and an
artificial viscosity term. The pressure term represents reversible work,
while the artificial viscosity term represents irreversible dissipation.

The artificial viscosity used in our simulations is given by \citep{ann05}
\begin{equation}
Q=
\begin{cases}
I_N\Delta x_{\rm min}\partial_iV^i(2\Delta x_{\rm min}
\partial_jV^j-0.3 c_s) & \text{ if $\partial_kV^k<0$},\\
0 & \text{ otherwise,}
\end{cases}
\label{eq:qdiss}
\end{equation}
where $\Delta x_{\rm min}$ is the minimum covariant zone length for the
particular grid point under consideration, and $c_{\rm s}$ is the
relativistic sound speed,
\begin{equation}
c_{\rm s}=\left[{\Gamma(\Gamma-1)P\over\Gamma P+(\Gamma-1)\rho}
\right]^{1/2},
\end{equation}
where $\rho$ is the fluid frame rest mass density.
The quantity $I_N$ is a normalized inertia multiplier,
\begin{equation}
I_N\equiv{1\over g_3^{1/2}}[E+W(\rho+P+\sqrt{3}Q+2P_B)],
\label{eq:inertia}
\end{equation}
where $P_B$ is the magnetic pressure in the fluid rest frame,
and $g_3$ is the 3-metric determinant.\footnote{This formulation of
the artificial viscosity is not covariant due to the presence of the coordinate velocity divergence in the definition.  We believe that this does not
significantly affect the results presented here, but in the future it would be
good to test this with a fully covariant formulation of artificial viscosity.}
Note that $I_N$ itself depends on $Q$. In the simulation this corresponds
to the $Q$ at the previous time-step.  We drop this term in the analysis
that follows, as it is difficult to reconstruct from the existing simulation
data.  We also tried solving equations (\ref{eq:qdiss}) and (\ref{eq:inertia})
simultaneously for $Q$, and the results of 
these two approaches were generally consistent within a few percent. However,
we occasionally found spurious numerical
artifacts (i.e. negative dissipations) when we tried solving the two equations
simultaneously. 

\begin{figure*}
\plotone{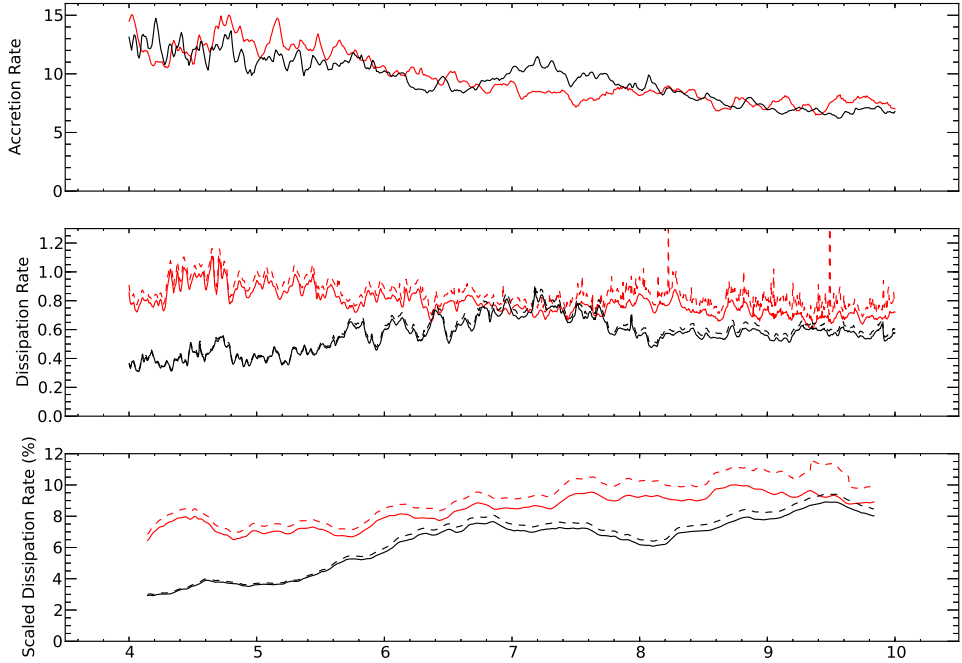}
\caption{Instantaneous accretion rate into the hole (top, in arbitrary units),
volume-integrated dissipation rate due to artificial viscosity (middle, in
arbitrary units), and volume-integrated dissipation
rate scaled by the instantaneous accretion energy $\dot{M}c^2$ into the
hole smoothed over a local time interval of $237M\simeq0.3\,t_{\rm orb}$
(bottom).  Black  curves refer to the $10^\circ$ simulation, while red  curves refer to the
$15^\circ$ simulation.  Dashed curves represent the full
volume-integrated dissipation rate in each simulation, while solid curves exclude the dissipation inside the coordinate radius of the
equatorial direct photon orbit as a crude estimate of how much of the
dissipation, if emitted locally, might reach an observer at infinity. }
\label{fig:mdotdissvst}
\end{figure*}

We integrated the artificial viscosity term $-Q\partial_i(WV^i)$ in
equation~(\ref{eq:internalenergy}) over the highest refinement grid portion
inside of $r=14M$ over all epochs to determine the total dissipation rate associated with the shocks.
We then time-averaged this total dissipation rate, and found that it amounts
to $6.1$ percent of the time-averaged rest mass energy accretion rate
$\dot{M}c^2$ onto the hole for the $10^\circ$ simulation, and $8.9$
percent for the $15^\circ$ simulation.  Figure~\ref{fig:mdotdissvst} shows
the instantaneous accretion rates, and instantaneous  dissipation.   It
is also useful to measure the instantaneous dissipation rate and
scale it with the instantaneous $\dot{M}c^2$ currently entering the hole.
We first smoothed $\dot{M}c^2$ and the dissipation rate over $237M\simeq0.3 \, t_{\rm orb}$
time intervals to smooth out high frequency fluctuations. As shown in the bottom panel, the scaled dissipation rate 
ranges from approximately 3 to 9 percent and 7 to 12 percent of the instantaneous $\dot{M}c^2$ for the $10^\circ$ and $15^\circ$
simulations, respectively.  Note that, apart from fluctuations, the accretion
rate into the hole generally declines by roughly a factor of two over long
time scales, while the dissipation rate shows no such obvious decline.  The
scaled dissipation rate therefore increases somewhat over time.

\begin{figure*}
\plottwo{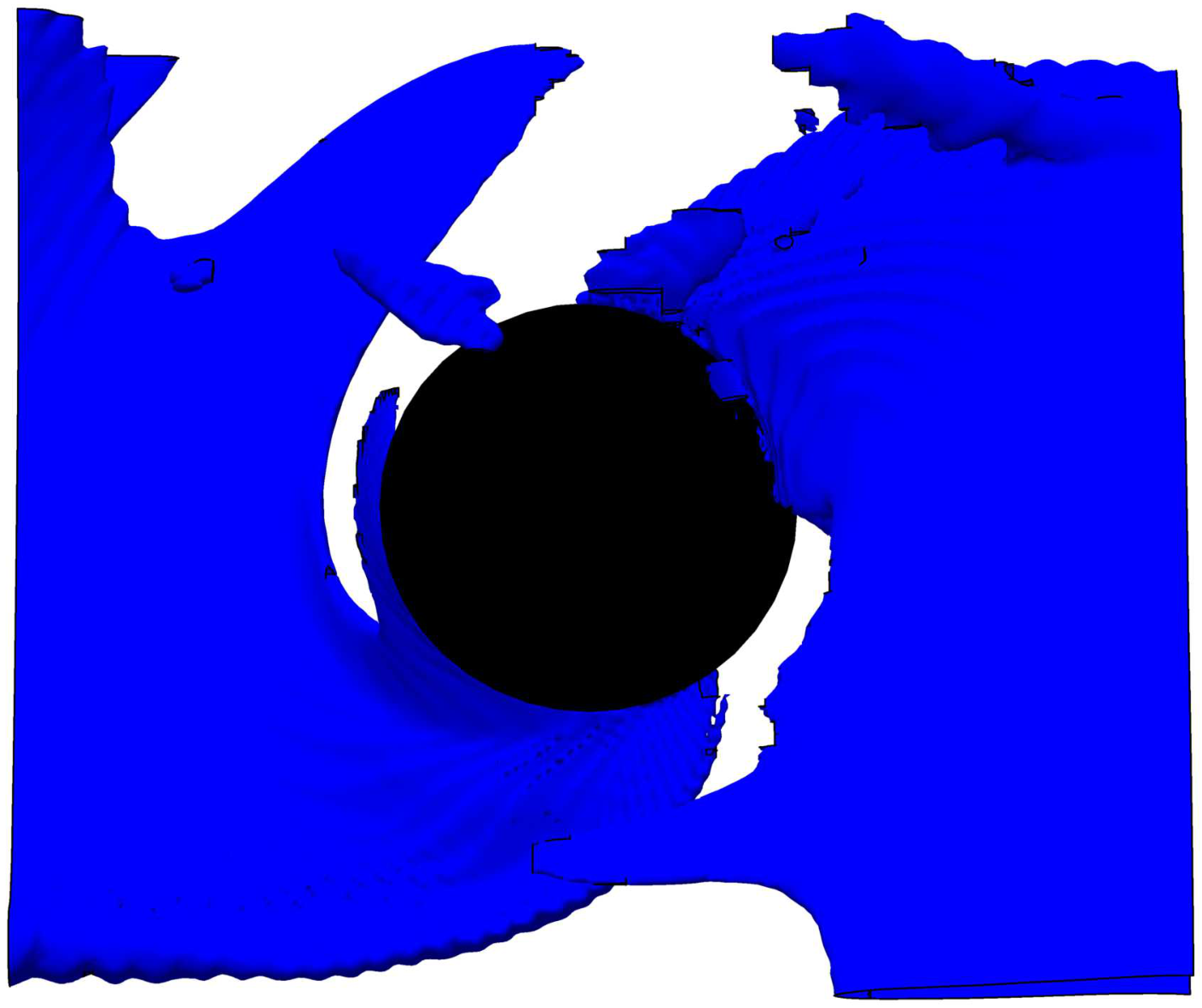}{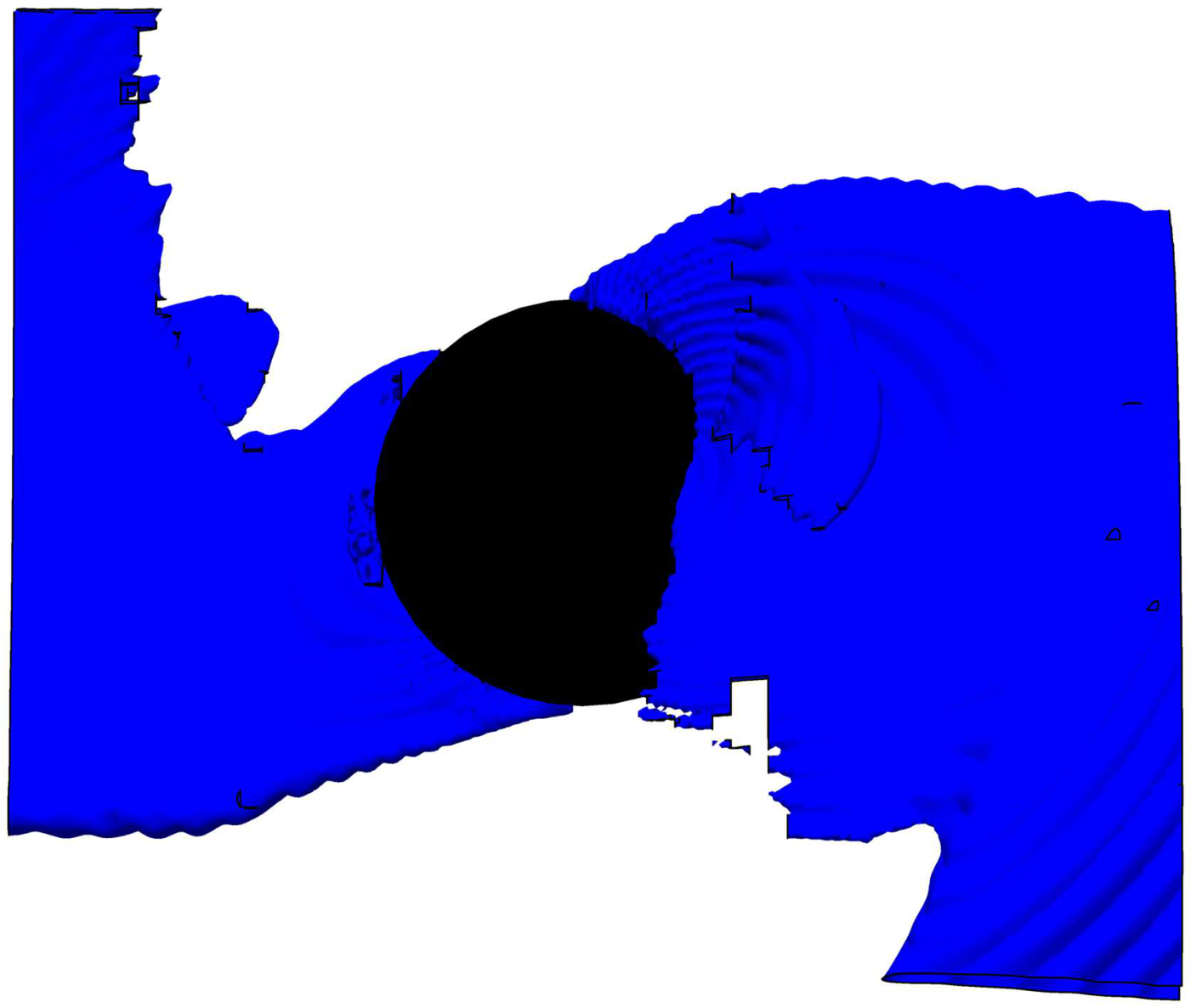}
\caption{Close-up (within a box 8M on a side) of the surface enclosing ninety
percent of the dissipation at epoch 9.95~$t_{\rm orb}$
for the $10^\circ$ simulation (left) and the $15^\circ$ simulation (right).
The viewing orientation is directly down the spin axis of the black hole,
and the flow in both simulations is orbiting counterclockwise in this
view.  The line of nodes between the black hole equatorial plane and the plane
of the initial torus is horizontal in the viewing angle of these figures.
As is particularly evident in the $15^\circ$ simulation on the right, two
shock surfaces have formed, one to the left and below the black hole equatorial
plane (i.e. behind the black hole in this viewing angle), and one to the
right and above the black hole equatorial plane (i.e. in front of the
black hole in this viewing angle).
}
\label{fig:dissipation}
\end{figure*}

That this dissipation is associated with shocks can be seen in
Figure~\ref{fig:dissipation}, which shows the spatial structure of surfaces
enclosing regions where ninety percent of the total dissipation occurs
in the $10^\circ$ and $15^\circ$ simulations. Note that in
Figure~\ref{fig:dissipation} we restrict the surface to a box that is $8M$
on each side.  The nonaxisymmetric structure
due to the presence of the large scale shocks is evident (see also Figures
\ref{fig:shockquantities910} and \ref{fig:shockquantities915} below).
For the purposes of this paper we take these 90 percent surfaces as a working
definition of shocks.
We note that our 90 percent surfaces are spatially co-located with the
entropy generation surfaces discussed in \citet{fra08,hen12}.

The spatial extent of the shocks is actually increasing slowly with time
throughout the course of both the 10$^\circ$ and 15$^\circ$ simulation.
Because of this, the shocks are able to dissipate a larger fraction of the
energy in the flow as time goes on.  In the case of the 10$^\circ$ simulation,
this more than compensates for the declining accretion rate from 4 to 7
orbits, so that the volume integrated dissipation rate increases over this
time period.  Later on, the increased spatial volume of the shocked regions
appears to just compensate for the declining accretion rate, and this is also
true throughout the 15$^\circ$ simulation, so that the dissipation rate is
roughly constant.

To address how much of this power escapes the vicinity of the
black hole requires an understanding of how that power might be emitted as
photons, and how many of those photons will reach an observer at infinity
\citep{dex11,dex12}.
As a crude estimate of this, Figure~\ref{fig:mdotdissvst} also shows the time
dependence of the
scaled volume-integrated dissipation rate in both simulations, but excluding
regions inside the coordinate radius of the direct equatorial photon orbit,
as it is likely that emission from inside this orbit will not reach infinity.
The dissipation rates outside the photon orbit are less than the previous
estimates of the total dissipation rates, but only slightly:
3 to 9 percent and 6 to 10 percent of the instantaneous
$\dot{M}c^2$ for the $10^\circ$ and $15^\circ$ simulations, respectively. 

These scaled dissipation rates should be compared with the binding energy per
unit rest
mass of the equatorial ISCO orbit, which is commonly used as an estimate of
the radiative efficiency for standard geometrically thin, optically thick
accretion disks.  This is approximately 16~percent for the $a/M=0.9$ black hole
spin assumed in our simulations.
Clearly the dissipation associated with shocks in accretion flows
with even the modest tilts considered here can be a substantial contributor
to the overall dissipation of accretion power.

\section{Dynamics Associated with the Shocks}
\label{sec:shockdynamics}

\begin{figure*}
\plottwo{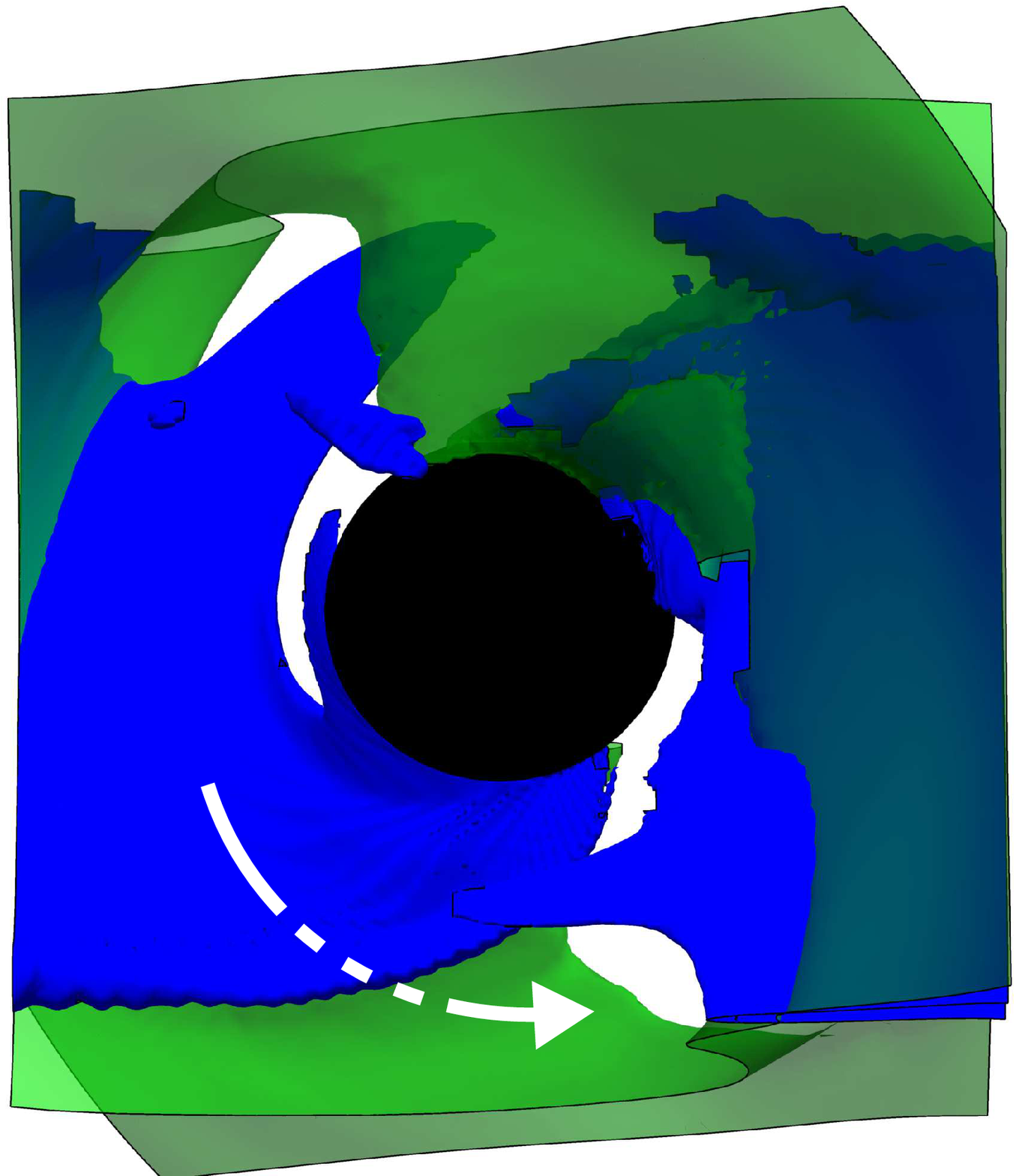}{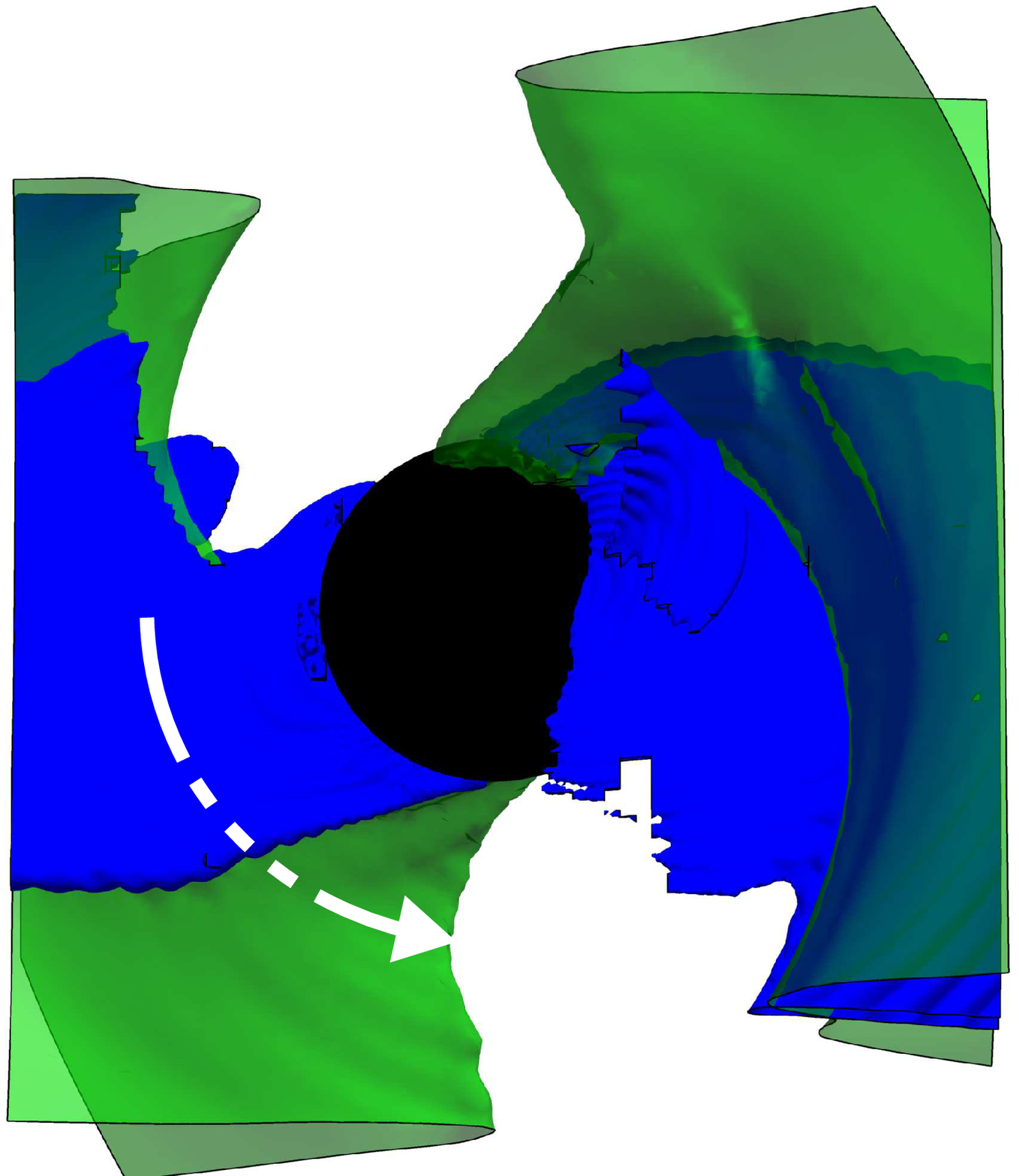}
\caption{Surface enclosing ninety percent of the dissipation (blue, identical
to that shown in Figure~\ref{fig:dissipation}) and a constant density
surface (green) for the $10^\circ$ simulation (left) and the $15^\circ$
simulation (right), both at epoch 9.95~$t_{\rm orb}$, within a box 8M on each
side.  The viewing orientation is exactly the same as in
Figure~\ref{fig:dissipation}), again looking directly down the spin axis of
the black hole, so that the sense of rotation of both the hole and the flow is
counterclockwise.  The arrows show the sense of a typical shock-crossing
trajectory
before (solid) it hits the left shock surface (blue), then is behind the shock
surface (dashed) in the compressed density region (green), and then emerges
from the compressed density region (solid).  The overdense regions (green)
clearly lie immediately downstream of the shock dissipation surfaces (blue).}
\label{fig:rhodiss}
\end{figure*}

In their analysis of radially extended coherent variability in tilted accretion
flows, \citet{hen12} showed that the two high density arms in the inner
parts of the accretion flow are located immediately downsteam of the two
standing shocks, and that structures associated with the variability
(transient density clumps and associated spiral acoustic waves) orbit
through the density arms.  This strongly suggests that the density arms
are regions of post-shock compression across which material continues to
move.  This differs from the interpretation of \citet{fra07}, which had
these density arms being actual fluid flow streams (``plunging streams"),
launched toward the black hole because geodesic orbits are unstable close
to the hole and/or the shocks extracted sufficient angular momentum from
the flow that streamlines turn toward the hole after crossing the shock
surfaces.  As illustrated in Figure~\ref{fig:rhodiss}, we likewise find that the
shock structures are clearly upstream of the overdense arms, and this is
consistent with both previous interpretations.

To understand the true nature of the high density arms, we here investigate
the fluid dynamics in their vicinity by examining the actual fluid trajectories.
We computed these trajectories using the simulation data
dumps of coordinate three-velocity $V^i\equiv u^i/u^t$ interpolated
in both space and time.
To advance a trajectory, we computed an interpolated velocity at the
current trajectory position and then used this to advance the trajectory
a constant time-step. We chose this time step to be smaller than the
nominal orbital time $2\pi(r^{3/2}+aM^{1/2})/M^{1/2}$ at the smallest
simulation radii, divided by the number of azimuthal grid zones in the
simulation.  As noted previously, our simulation data
dumps are separated by $\simeq3.955M$ in Kerr-Schild coordinate time,
corresponding to $\simeq1/7$ of an orbital period for the equatorial ISCO
orbit at $r\simeq2.32M$.  Our trajectories should therefore be reasonably
accurate even near the black hole, provided we do not integrate them for
much longer than an orbital period, and will be increasingly accurate
further away from the hole.

We selected grid points inside one of the density arms, and then integrated
the trajectories both forward and backward in time.  For
simplicity, we first restricted the starting grid points to all lie on one
radial shell ($r=2.03M$).
Then, in order to get points inside one arm, we chose points in the upper
hemisphere where the density exceeded the density of the contour chosen
to illustrate the arms shown in Figure~\ref{fig:rhodiss}.
For clarity of visualization, we then used only every fourth trajectory.

\begin{figure}
\plotone{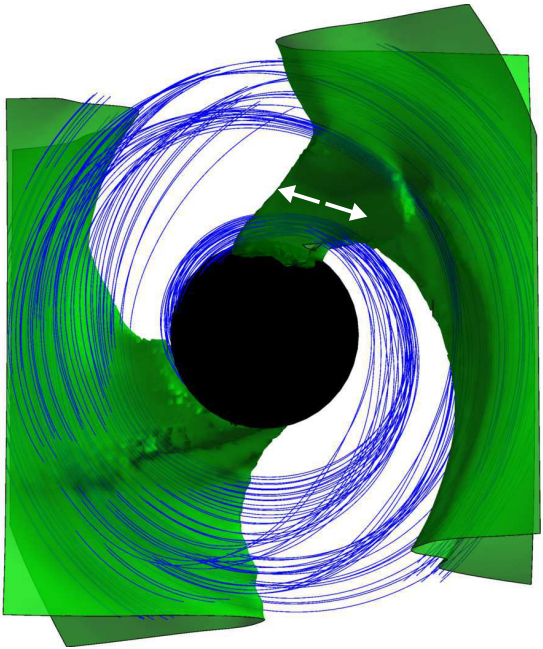}
\caption{The same constant density surface (green) as shown in the right hand
panel of Figure~\ref{fig:rhodiss}, with the same viewing orientation,
together with selected fluid element trajectories
(blue curves) for the $15^\circ$ simulation at epoch 9.95~$t_{\rm orb}$.
These trajectories were calculated by starting forward and backward
integrations at points within one of the shock dissipation surfaces,
as illustrated by the double arrow in the figure.  Fluid flows along
these trajectories in a counterclockwise sense.
}
\label{fig:rhotrajectories}
\end{figure}

Figure~\ref{fig:rhotrajectories} shows the resulting trajectories for the
$15^\circ$ simulation near epoch $9.95t_{\rm orb}$, that were
integrated both forward and backward in time starting from inside the density
arm in the upper right of the figure.  All the fluid trajectories clearly pass
{\it through} the high
density arm, which therefore must represent a post-shock compression region
rather than a flow stream.  The trajectories themselves still represent
plunging orbits, in that they spiral into the hole without going around many
times, and this is consistent with the fact that spheroidal test particle
orbits (non-equatorial geodesics with constant coordinate radius) are
unstable in this region \citep{fra07}.  Note, however, that these particular
fluid trajectories are not themselves geodesic orbits at large radii, as
pressure forces keep them moving above (or below) the disk midplane.

\section{Strength of Shocks}
\label{sec:shockstrength}

As noted in the introduction, \citet{fra08} claimed that the strengths
of the pairs of shocks formed in these tilted accretion flow simulations
were quite modest, but we now show that in fact they are quite strong.
One way to do this is to directly measure the upstream Mach number of the
shock at a particular point along its surface, but this is nontrivial because
of the complicated geometry of both the shock and the
tilted accretion flow.  We have instead adopted the strategy of computing
actual fluid trajectories in the vicinity of the shock surfaces, interpolating
fluid rest frame pressure and rest mass density in space and time
onto those trajectories, and then measuring
the jumps in these quantities that are caused by the shock.
The ideal hydrodynamic relativistic shock jump conditions then imply that
the upstream spatial three velocity component normal to the
local plane of the shock in the shock rest frame is given by
(e.g. \citealt{ani89}):
\begin{equation}
v_+=\left[{P_--P_+\over\rho_+^2(\tau_+-\tau_-)+P_--P_+}\right]^{1/2},
\end{equation}
where the subscripts $+$ and $-$ refer to upstream and downstream, respectively.
The quantity $\tau$ is the dynamical volume,
\begin{equation}
\tau\equiv{P+\rho+e\over\rho^2}.
\end{equation}
The upstream Mach number is then just ${\cal M}_+\equiv v_+/c_{{\rm s}+}$.

These formulas assume that magnetic pressure and tension are completely
negligible in the shock dynamics.  We enforced this by only using 
trajectories along which the local fluid rest frame magnetic pressure was
less than one tenth of the local fluid rest frame gas pressure everywhere.

\begin{figure*}[htbp]
\centering
\begin{minipage}[t]{0.40\linewidth}
\centering
\includegraphics[width=\linewidth]{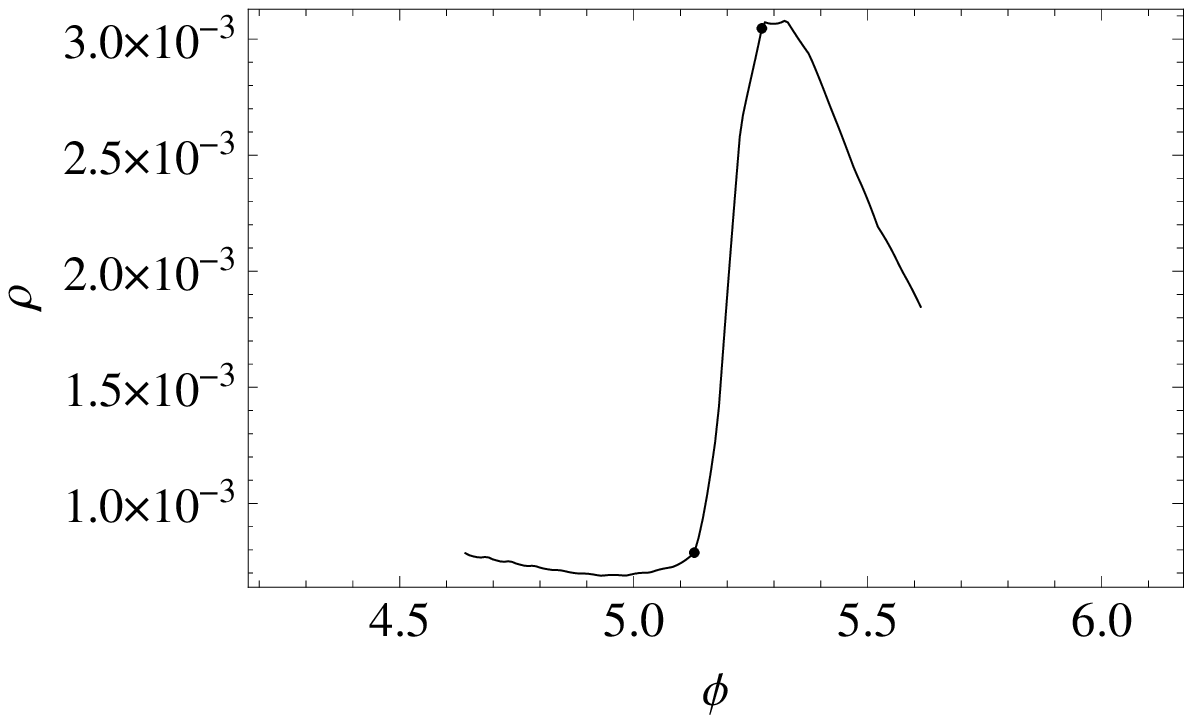}
\end{minipage}\\
\begin{minipage}[t]{0.40\linewidth}
\centering
\includegraphics[width=\linewidth]{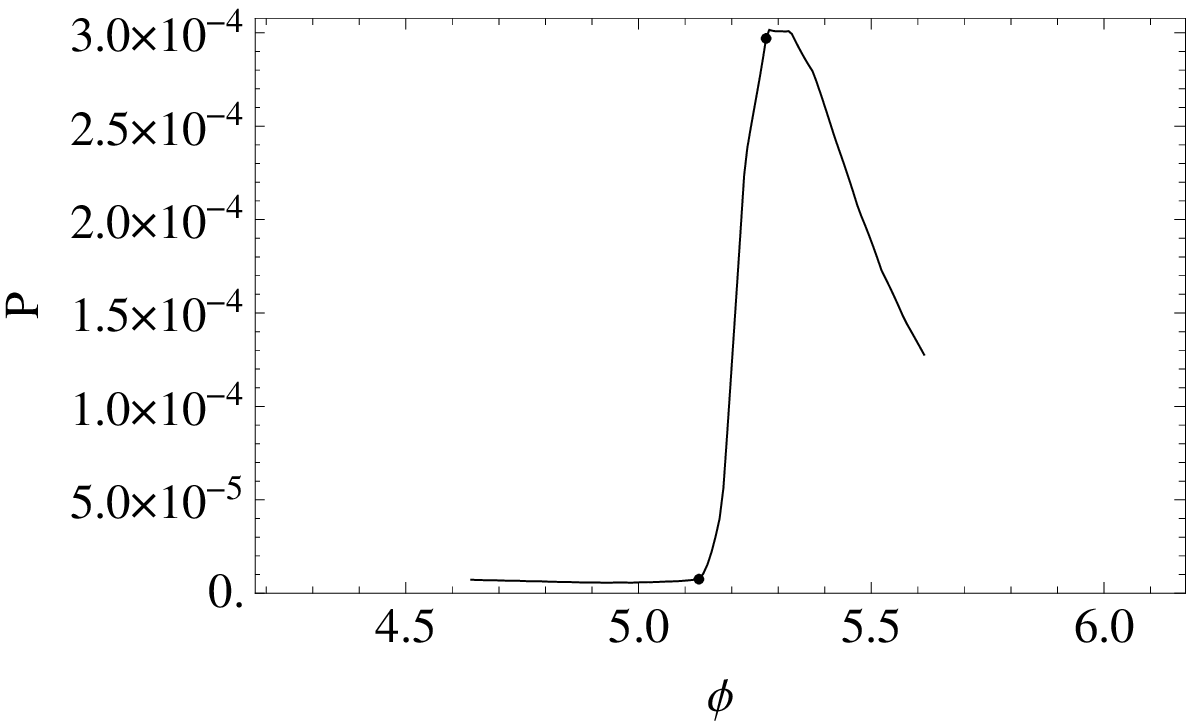}
\end{minipage}\\
\begin{minipage}[t]{0.40\linewidth}
\centering
\includegraphics[width=\linewidth]{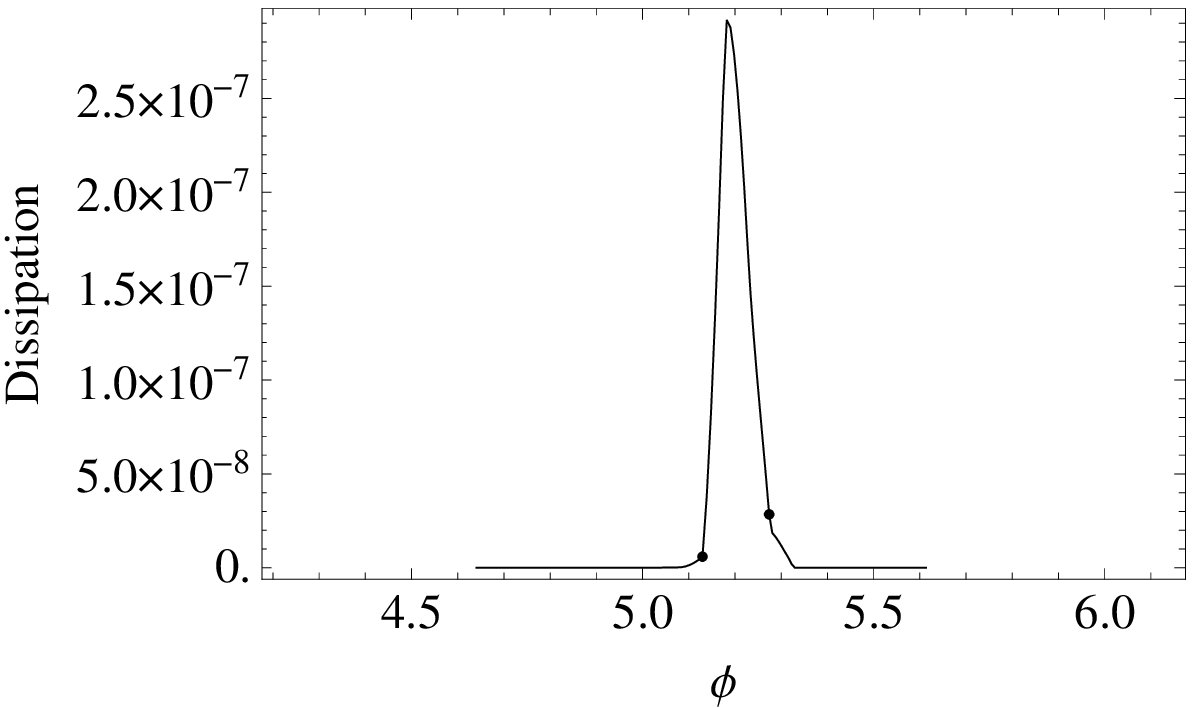}
\end{minipage}%
\caption{Density $\rho$, pressure $P$, and dissipation rate per unit volume,
all in code units, along one particular trajectory as it crosses the shock
surface at epoch 9.95~$t_{\rm orb}$ in the $15^\circ$ tilt simulation.  Dots
show the locations where pre- and post-shock quantities were measured.}
\label{fig:shockjumps}
\end{figure*}

We measured jumps in pressure and density along the fluid trajectories
as they cross the shock surfaces. For each point on the 
shock surfaces shown in Figure~\ref{fig:rhodiss} we calculated
short trajectory segments
going forward and backward in time. To ensure our measurements remain local, we restricted both
the forward- and reverse-time trajectory segments to ten azimuthal zones
(twenty for the combined trajectory). We experimented with the length of trajectory segments calculated
and found that increasing the number of azimuthal zones slightly (from ten to fifteen) does not affect our results. 
In order to precisely measure the pre- and post-shock pressures and densities, we also measured the artificial
viscosity dissipation rate per unit volume [$-Q\partial_i(WV^i)$ in equation
(\ref{eq:internalenergy})] along the trajectories.  This generally
exhibits a very sharp maximum near the shock surfaces. 
Thus, we chose to measure pre- and post-shock fluid quantities
at positions on either side of the dissipation
peak where the dissipation rate dropped to below eleven percent of its maximum. Any trajectory segment for which this did 
not occur was excluded. We experimented with different percentage thresholds (i.e. five and twenty percent)
and found that the maximum upstream Mach numbers, Lorentz factors and
compression ratios changed by at most five percent. (Interestingly, measurements
for individual trajectories were not as robust. For example, we took trajectories
corresponding to the maximum Mach number for a given threshold and measured the changes in the 
quantities for each individual trajectory in this set.  We found that the shock 
quantities could change by as much as twenty percent. In fact, trajectories
corresponding to maximum shock quantities for our nominal threshold 
could be rejected for other choices of threshold. While these details may merit further investigation, they are not particularly 
important for the purposes of obtaining a rough estimate of overall shock strength.)  As an independent
verification of whether these provided
good pre- and post-shock measurements, we also checked how well they satisfied
the Taub adiabat (e.g. \citealt{ani89}), which may be written
\begin{equation}
{(P_--P_+)(\tau_-+\tau_+)\over\rho_-^2\tau_-^2-\rho_+^2\tau_+^2}=1.
\label{eq:taub}
\end{equation}
We only considered trajectories for which the difference between the
left hand side and unity was less than five percent over the shock surfaces.
Figure~\ref{fig:shockjumps} shows an example of fluid
quantities measured along one particular trajectory that we considered
acceptable at epoch 9.95~$t_{\rm orb}$ in the $15^\circ$ tilt simulation.

\begin{figure*}[htbp]\centering
\begin{minipage}[t]{0.40\linewidth}
\centering
\includegraphics[width=\linewidth]{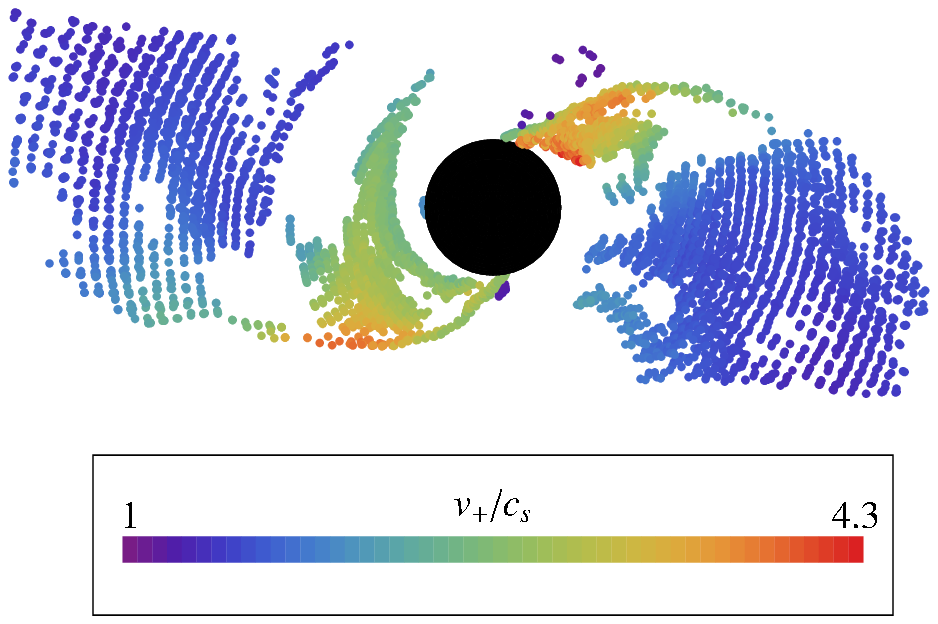}
\end{minipage}%
\begin{minipage}[t]{0.40\linewidth}
\centering
\includegraphics[width=\linewidth]{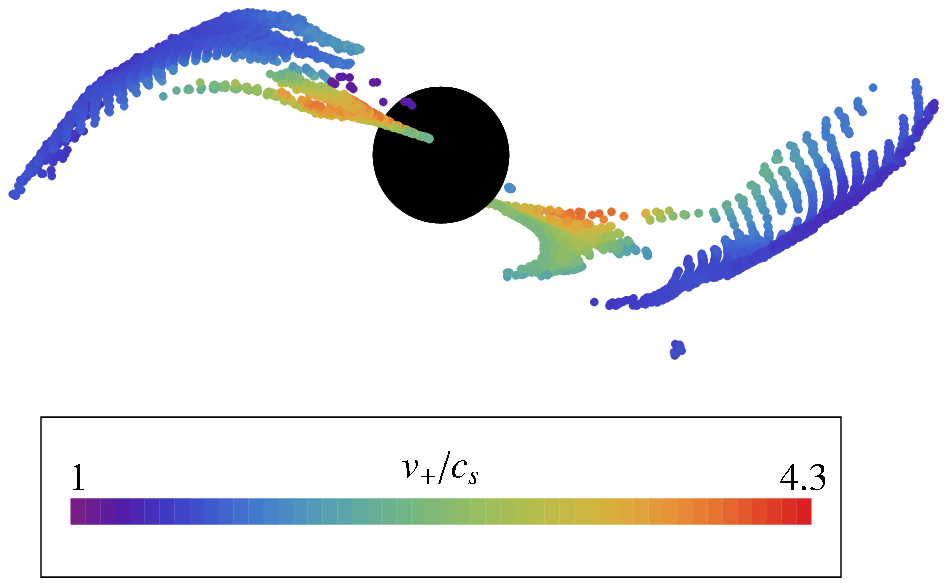}
\end{minipage}\\
\begin{minipage}[t]{0.40\linewidth}
\centering
\includegraphics[width=\linewidth]{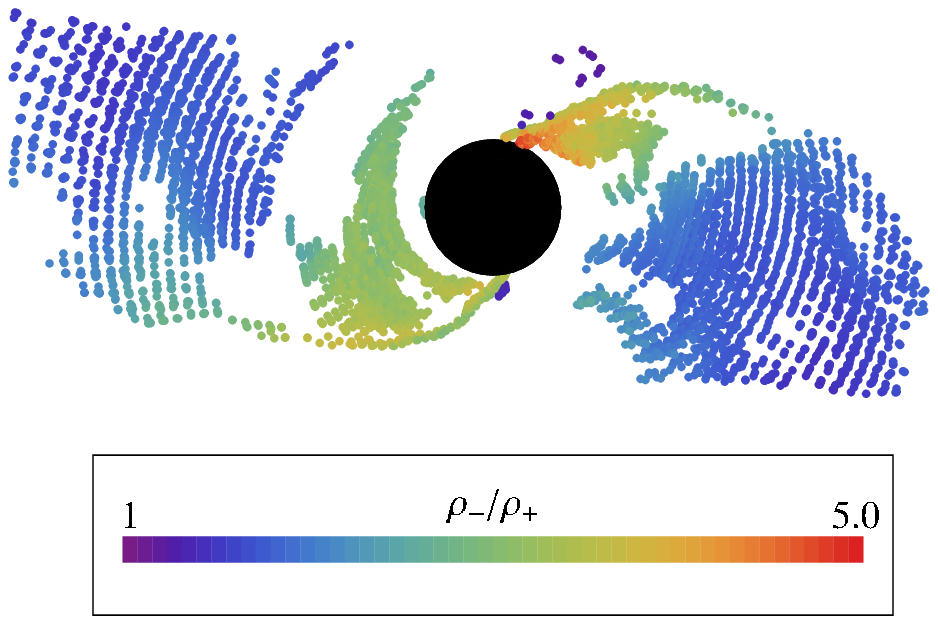}
\end{minipage}%
\begin{minipage}[t]{0.40\linewidth}\centering
\includegraphics[width=\linewidth]{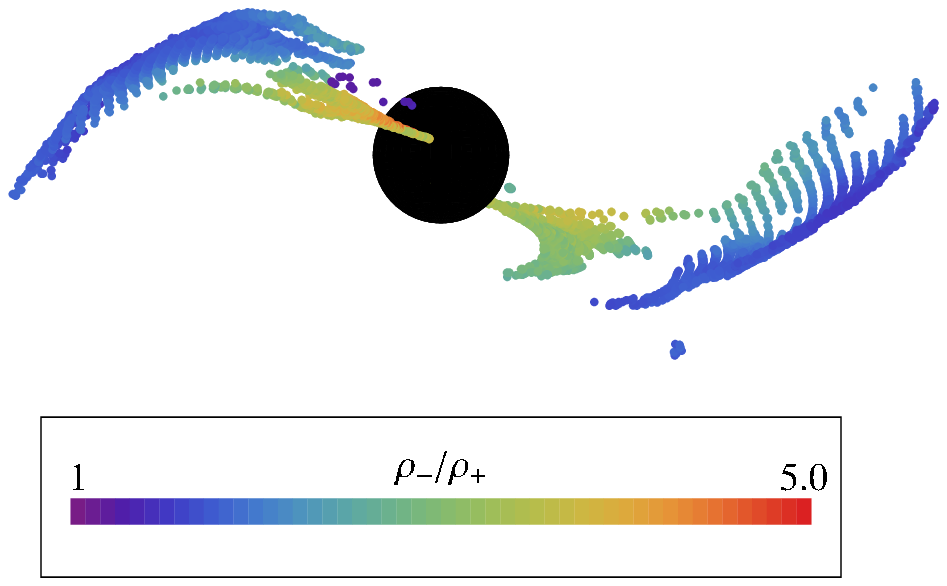}\\
\end{minipage}\\
\begin{minipage}[t]{0.40\linewidth}
\centering
\includegraphics[width=\linewidth]{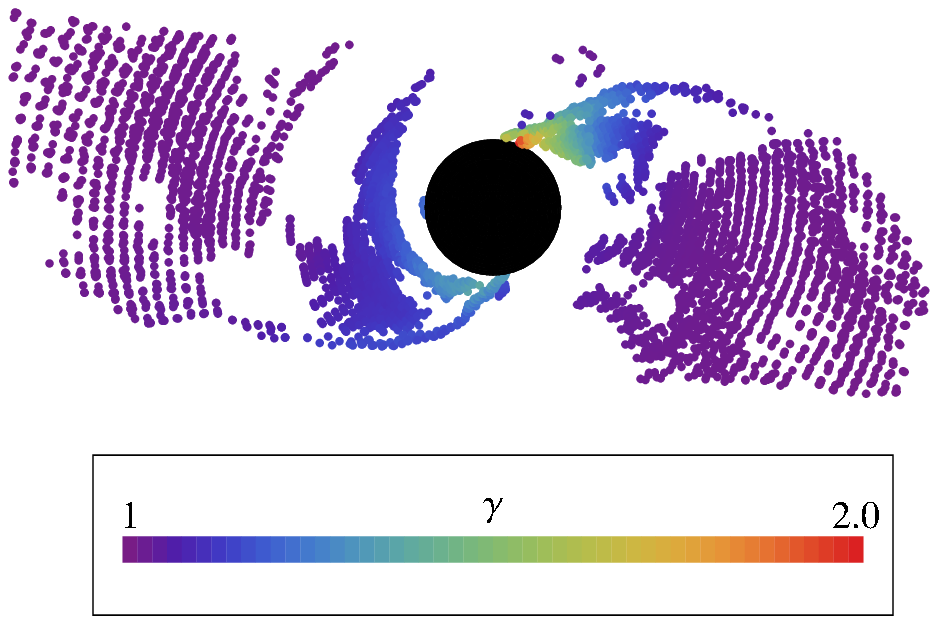}
\end{minipage}
\begin{minipage}[t]{0.40\linewidth}
\centering
\includegraphics[width=\linewidth]{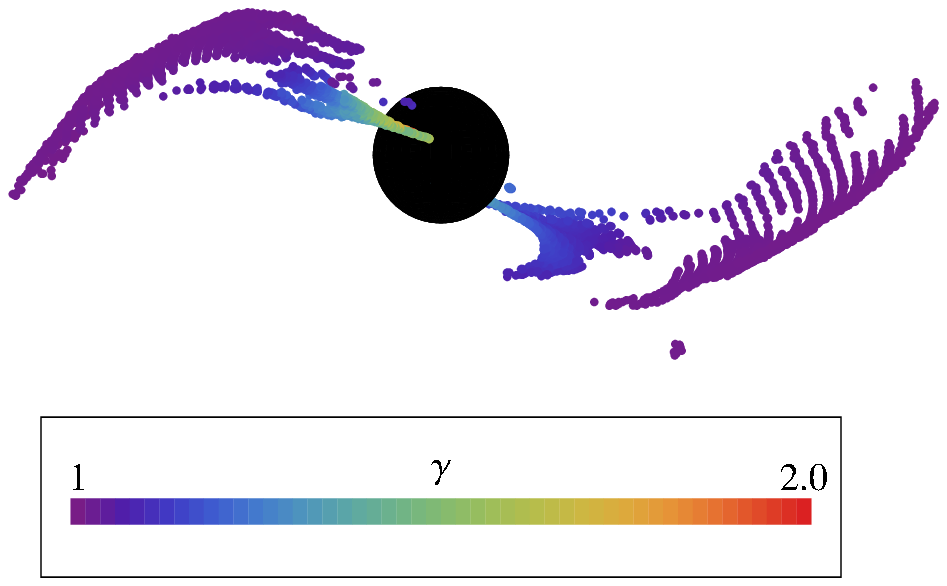}
\end{minipage}
\caption{Upstream Mach number ${\cal M}_+=v_+/c_{{\rm s}+}$ (top row),
Lorentz factor $\gamma(v_+)\equiv(1-v_+^2)^{-1/2}$ of the upstream
three-velocity component $v_+$ normal to the shock surface in the local
shock rest frame (middle row), and compression ratio $\rho_-/\rho_+$
(bottom row) for the $10^\circ$ tilt simulation 0910h at epoch
$9.95t_{\rm orb}$.  The left column figures are viewed from directly above
the black hole spin axis, with the line of nodes between the black
hole equatorial plane and the plane of the initial torus oriented horizontally
(the same viewing geometry as in
Figures~\ref{fig:dissipation}-\ref{fig:rhotrajectories}).
The sense of rotation of both the black hole and the accretion flow is
counterclockwise.
The right column figures are viewed with the black hole spin axis pointing
upward in the plane of the page, and viewed from a direction in the equatorial
plane that corresponds to being above the position of the black hole
in the left hand column
of figures.  The line of nodes between the black hole equatorial plane and
the plane of the initial torus is still horizontal and in the plane of the page
in this view.  The figures are
to scale (treating the spatial Kerr-Schild coordinates as flat space spherical
polar coordinates for constructing these figures), with the filled black circle
indicating the location of the event horizon at $r\simeq1.44M$.
The horizontal extent of the figures is approximately $20M$.}
\label{fig:shockquantities910}
\end{figure*}

\begin{figure*}[htbp]
\centering
\begin{minipage}[t]{0.40\linewidth}
\centering
\includegraphics[width=\linewidth]{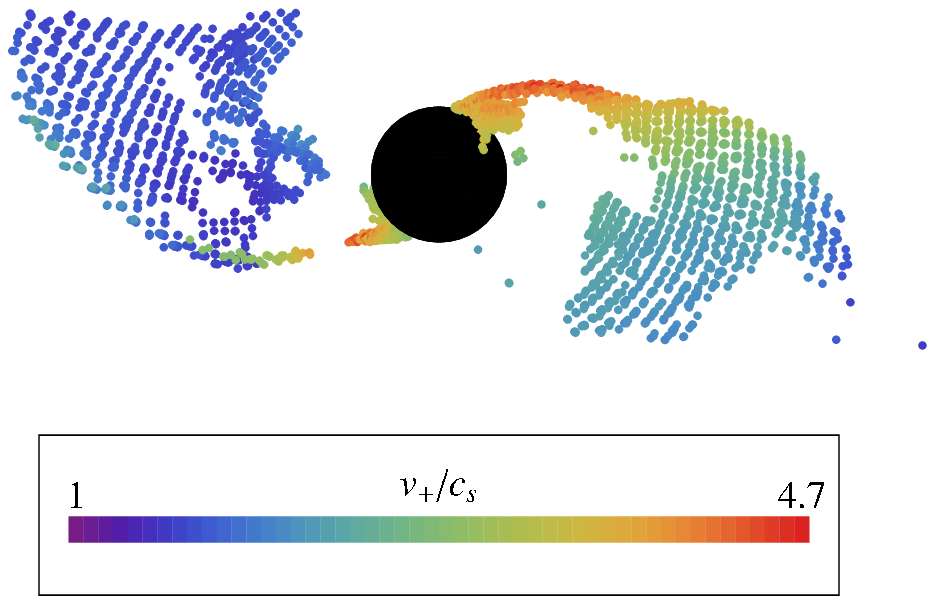}
\end{minipage}%
\begin{minipage}[t]{0.40\linewidth}
\centering
\includegraphics[width=\linewidth]{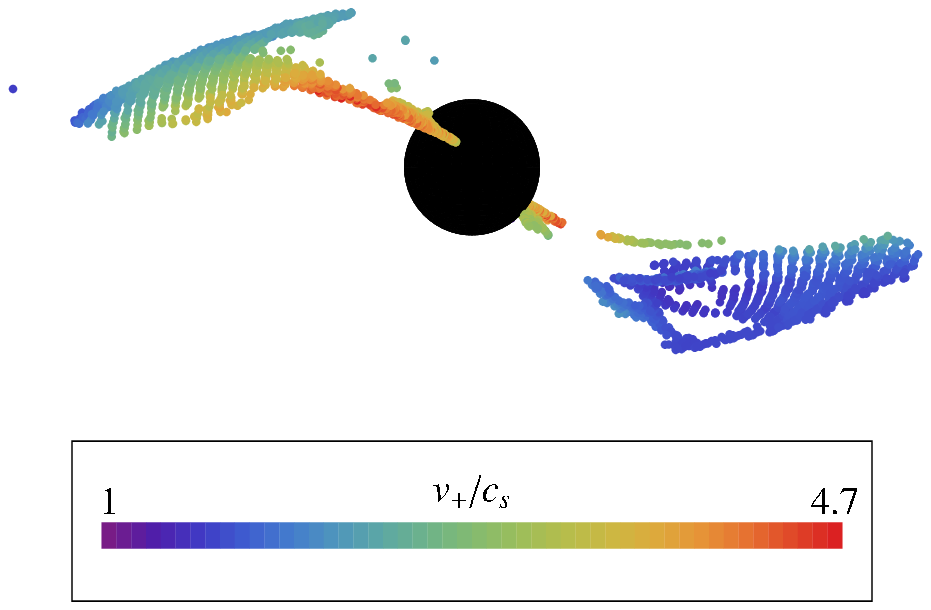}
\end{minipage}\\
\begin{minipage}[t]{0.40\linewidth}
\centering
\includegraphics[width=\linewidth]{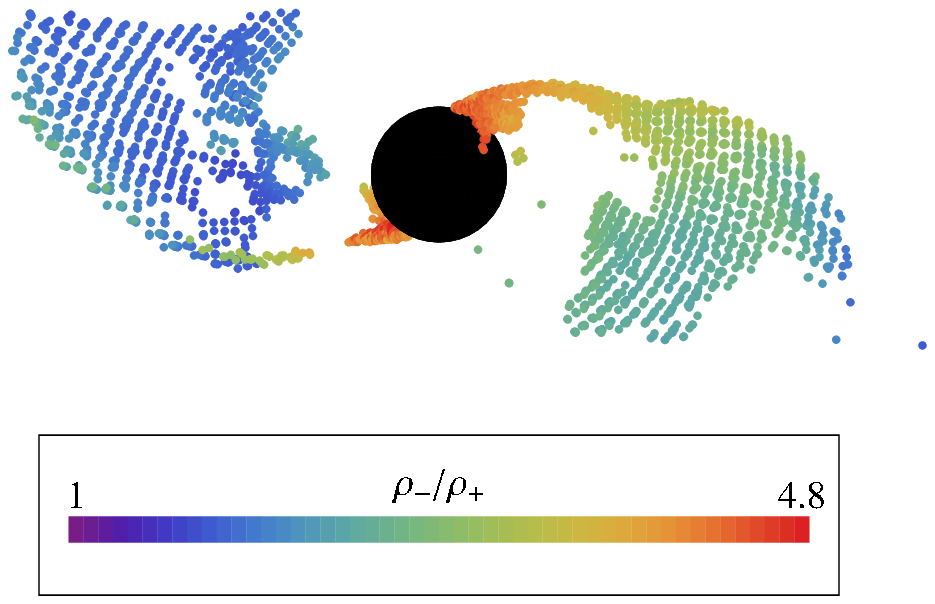}
\end{minipage}%
\begin{minipage}[t]{0.40\linewidth}
\centering
\includegraphics[width=\linewidth]{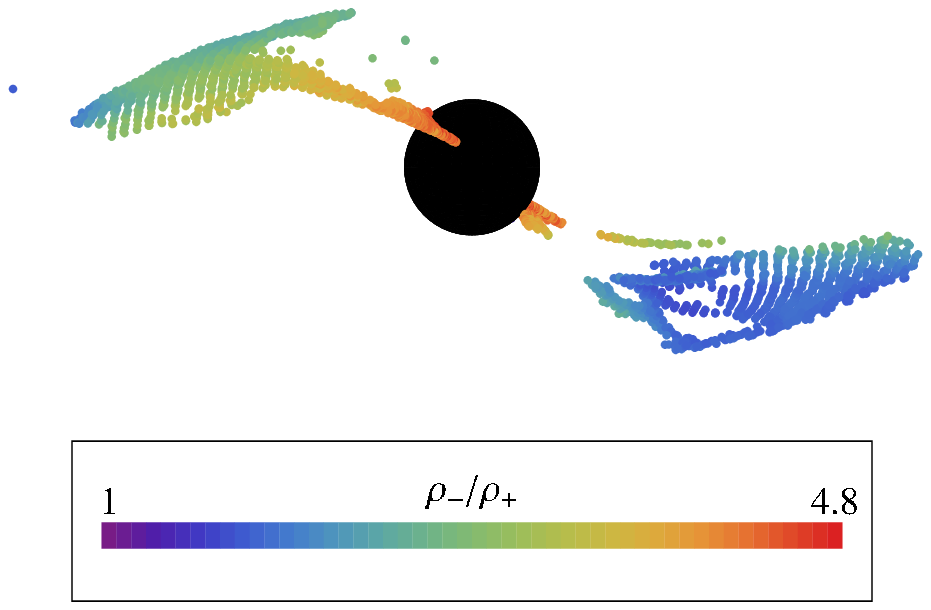}\\
\end{minipage}\\
\begin{minipage}[t]{0.40\linewidth}
\centering
\includegraphics[width=\linewidth]{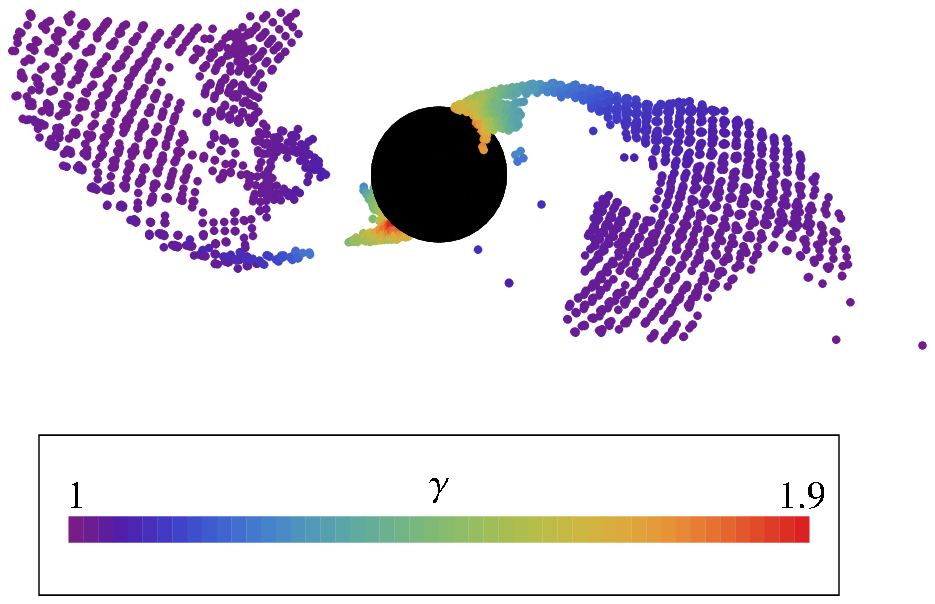}
\end{minipage}%
\begin{minipage}[t]{0.40\linewidth}
\centering
\includegraphics[width=\linewidth]{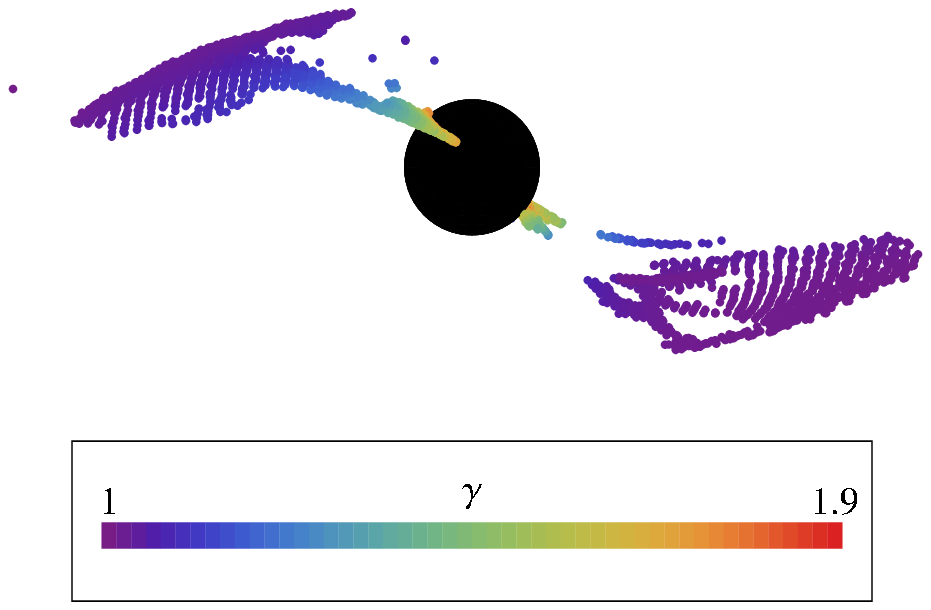}
\end{minipage}
\caption{Same as Figure~\ref{fig:shockquantities910}, only for the  $15^\circ$
tilt simulation 0915h at epoch $9.95t_{\rm orb}$.}
\label{fig:shockquantities915}
\end{figure*}

Figures~\ref{fig:shockquantities910} and \ref{fig:shockquantities915} show
the distribution of upstream Mach numbers, Lorentz factors, and
density compression ratios across the shock surfaces
at epoch $9.95~t_{\rm orb}\simeq7871M$ in the $10^\circ$ and $15^\circ$ tilted
simulations, respectively.  The shock surfaces in these figures have
projected radial extents $\lesssim10M$, consistent with the radial extent
of significant entropy generation measured in these simulations
\citep{fra08,hen12}.  The shocks are mildly relativistic, with
upstream Lorentz factors of at most 2 in the $10^\circ$ tilt simulation
and 1.9 in the $15^{\circ}$ tilt simulation at this particular epoch, and even
these Lorentz factors are only achieved very close to the black hole ($r=1.5M$
and $1.6M$, respectively, compared to the horizon radius of $1.44M$).  However,
the shocks are nevertheless strong, with upstream Mach numbers ${\cal M_+}$ as
high as 4.3 at approximately $r=2.3M$ and 4.7 at approximately $3.4M$ for the
$10^\circ$ and $15^\circ$ tilted simulations, respectively, at this particular
epoch.
Density compression ratios are also strong, as high as $4.8$ and $5$ for $10^\circ$ and $15^\circ$, respectively, at this particular epoch.
This is somewhat higher than the maximum value of 4 expected for a
non-relativistic shock, and these high compression ratios are occurring near
the black hole where the Lorentz factors of the flow through the shock are
somewhat higher than unity.  The highest Mach numbers are actually achieved
further out in radius, where the compression ratios are 4.3 and 4.1 for the
$10^\circ$ and $15^\circ$ simulations, respectively.

Table 1 summarizes the properties of the shocks at positions of maximal
upstream Mach number at this epoch as well as three earlier epochs, in both
simulations.  Also listed are the locations and values where the shocks
achieve maximal upstream Lorentz factor and compression ratio.
The maximal Mach numbers vary considerably with epoch, ranging from 3 to 4.6
for 10$^{\circ}$ and 4.7 to 6 for 15$^{\circ}$.   The maximal upstream
Lorentz factors never exceed 2 for all the tilts and epochs considered
here.   Material is also strongly compressed with the compression ratio
exceeding 4 for 2 of the 4 epochs for 10$^{\circ}$ and all 4 epochs for
15$^{\circ}$.  The maximum Mach number generally occurs further out in radius
from the point of maximal compression.  Although the shock properties vary
over time, the shocks are consistently strong and only mildy relativistic.

\begin{deluxetable*}{cccccccccccc}
\tablecaption{Shock Properties at Various Epochs and Locations}
\tablehead{\colhead{Tilt} & \colhead{Epoch} &
\multicolumn{4}{c}{Maximum Upstream Mach Number} &
\multicolumn{2}{c}{Maximum Upstream Lorentz Factor} &
\multicolumn{2}{c}{Maximum Compression Ratio}\\
\colhead{($^{\circ}$)} & \colhead{(t$_{\rm orb}$)} &
\colhead{$(r/M,\vartheta,\varphi)$\tablenotemark{a}} &
\colhead{$v_+/c_{\rm s}$} & \colhead{$\gamma(v_+)$} &
\colhead{$\rho_-/\rho_+$} &
\colhead{$(r/M,\vartheta,\varphi)$\tablenotemark{a}} &
\colhead{$\gamma(v_+)$} &
\colhead{$(r/M,\vartheta,\varphi)$\tablenotemark{a}} &
\colhead{$\rho_-/\rho_+$} }
\startdata
10 & 4.05 & (2.5,2.1,3.1) & 3.0 & 1.1 & 3.3 & (1.6,1.6,1.7) & 1.1 & (2.3,1.8,3.8) & 3.4 &  \\
10 & 6.00 & (3.6,1.3,6.0) & 3.8 & 1.1 & 3.6 & (1.5,1.8,4.0) & 1.5 & (1.5,1.8,4.0) & 3.9 &  \\
10 & 8.00 & (4.7,1.7,2.7) & 4.6 & 1.2 & 3.9 & (1.5,0.81,6.0) & 1.5 & (3.2,1.8,3.1) & 4.1 &  \\
10 & 9.95 & (2.3,0.99,5.1) & 4.3 & 1.5 & 4.3 & (1.5,1.1,5.9) & 2.0 & (1.5,1.1,5.9) & 5.0 &  \\
15 & 4.05 & (2.8,1.9,2.9) & 4.7 & 1.3 & 4.1 & (1.6,2.3,1.9) & 1.4 & (2.3,1.9,3.1) & 4.4 &  \\
15 & 6.00 & (2.4,1.2,6.1) & 6.0 & 1.5 & 4.5 & (1.5,0.92,6.1) & 1.6 & (2.1,1.2,6.2) & 4.7 &  \\
15 & 8.00 & (4.1,2.1,1.9) & 5.0 & 1.2 & 4.0 & (1.5,2.3,2.6) & 1.9 & (1.7,2.3,2.6) & 4.9 &  \\
15 & 9.95 & (3.5,1.0,5.2) & 4.7 & 1.2 & 3.9 & (1.6,2.2,2.3) & 1.9 & (1.6,2.2,2.3) & 4.8 \\
 &  \\
\enddata
\tablenotetext{a}{Locations in tilted Kerr-Schild coordinates}
\end{deluxetable*}

Our previous claim that the shocks were quite weak was based in
part on the fact that there exist regions in the flow which are strongly
magnetized (low plasma beta), and visually these regions appeared to be
associated with the post-shock regions of the flow \citep{fra08}.
However, our more detailed, quantitative analysis here shows that much of
the shock surfaces are actually associated with {\it weakly} magnetized fluid
trajectories. Our previous argument was therefore incorrect,
and the shocks are in fact strong. In this paper, we focused on weakly
magnetized trajectories, 
purposely excluding trajectories passing through strongly magnetized regions.  

\section{Conclusions}
\label{sec:conclusions}

In general, there is no reason to expect the angular momentum of the fuel
source in a black hole accretion disk to have any causal connection to the
spin of the hole itself, unless there has been sufficient time for alignment
to take place.  This suggests that tilted accretion flows could be
quite common in nature, particularly in the case of supermassive black holes
which frequently receive new sources of fuel.  Two mechanisms have been
proposed that can plausibly align a black hole and a tilted accretion disk.
A geometrically thin disk could
be aligned by the Bardeen-Petterson effect \citep{bar75,sor13b}.
More recently, an
MHD alignment mechanism associated with magnetically arrested accretion was
found to be effective in aligning geometrically thick disks \citep{mck13}.
Nonetheless, there are GRMHD simulations of geometrically thick flows where
the disk remains misaligned with the spin of the hole \citep{fra07}.

Such simulations of hot tilted flows exhibit striking qualitative differences
in the innermost regions from untilted flows:  notably, a pair of standing
shocks and a pair of high density arms.  Fluid trajectories pass through the
high density arms, which are therefore regions of post-shock compression,
rather than streams of material plunging into the
hole as claimed by \citet{fra07}.  The shocks themselves are quite strong,
though only mildly relativistic, with upstream Mach numbers as high as
$\simeq4.6-6$ for tilt angles of $10^\circ-15^\circ$.  Dissipation within
the shocks amounts to a substantial fraction ($3-12$ percent) of $\dot{M}c^2$.
A steady-state has not yet been achieved in the simulations analyzed in
this paper, and the dissipated fraction of $\dot{M}c^2$ actually slowly
increases with time.  This is because the shocked regions expand outward
and dissipate larger fractions of the accretion power.

It is natural to ask how the shocks would affect the observed properties of
the accretion flow.  
Synthetic images, spectra and variability of simulated tilted accretion flows
produced by post-processed general relativistic ray tracing calculations have
been produced in recent years using a number of emission models
\citep{dex11, dex12}.  In the context of the Galactic Center source Sgr~A$^*$,
the emission was assumed to be thermal synchrotron with a spatial and
temporal electron temperature distribution that is in fixed proportion to the
ion temperature measured from the simulations \citep{dex12}.  Synthetic
images under these assumptions are strongly dominated by the shock structures,
particularly when viewed face-on.  Moreover, the electron temperatures get
high enough that the spectra can reproduce the observed near infrared emission
from this source.  That the shocks are playing such a dominant role may in
part be because of the assumed emission model:  the internal energy evolution
scheme of the simulations means that the only heating that occurs is near
shocks, and so that is where the electron temperatures will be high.  However,
the fact that we find that the dissipation produced by the shocks is a
substantial fraction of $\dot{M}c^2$ implies that the shocks
should be a very important contributor to the emission from tilted accretion
flows, regardless of the assumed emission mechanism.  For example, \citet{dex09}
estimate that of order ten percent of $\dot{M}c^2$ is lost by magnetic
reconnection in an untilted simulation completed using the same,
non-energy conserving code that we use here.  Shock dissipation is therefore
likely to be comparable to turbulent dissipation in tilted accretion flows.

Particularly in the collisionless conditions of the Sgr~A$^*$ flow, the
presence of standing shocks could also lead to particle acceleration and
nonthermal electron energy distributions.  We find that the shocks are strong,
but not very relativistic, with compression ratios of at most 5.
Standard first order Fermi acceleration models would therefore predict
synchrotron surface brightness spectral indices
$(I_\nu\propto\nu^{\alpha})$ as flat as $\alpha=-0.4$.

\acknowledgements
We thank the referee for comments that significantly improved this paper,
and Jason Dexter for very useful conversations and comments.
This work was supported in part by National Science Foundation grant
AST-0707624.  In addition, AG acknowledges support from the Goldwater
Foundation and PCF acknowledges support from the National Science Foundation
under grants AST-0807385 and PHY11-25915.


\begin{thebibliography}{}
\bibitem[Anile(1989)]{ani89}Anile, A. M. 1989, Relativistic Fluids and
Magneto-Fluids (Cambridge:  Cambridge University Press)
\bibitem[Anninos, Fragile \& Salmonson(2005)]{ann05}Anninos, P., Fragile,
P. C., \& Salmonson, J. D. 2005, ApJ, 635, 723
\bibitem[Balbus \& Hawley(1991)]{bal91}Balbus, S. A., \& Hawley, J. F. 1991,
ApJ, 376, 214
\bibitem[Balbus \& Hawley(1998)]{bal98}Balbus, S. A., \& Hawley, J. F. 1998,
Rev. Mod. Phys., 70, 1
\bibitem[Bardeen \& Petterson(1975)]{bar75}Bardeen, J. M., \& Petterson,
J. A.  1975, ApJ, 195, L65
\bibitem[Beckwith et al.(2006)]{bec06}Beckwith, K., Hawley, J. F., \& Krolik,
J. H. 2006, eprint (astro-ph/0605295)
\bibitem[Beckwith et al.(2008)]{bec08}Beckwith, K., Hawley, J. F., \& Krolik,
J. H. 2008, MNRAS, 390, 21
\bibitem[Cuadra et al.(2006)]{cua06}Cuadra, J., Nayakshin, S., Springel,
V., \& Di Matteo, T. 2006, MNRAS, 366, 358
\bibitem[De Villiers, Hawley \& Krolik(2003)]{dev03}De Villiers, J.-P.,
Hawley, J. F., \& Krolik, J. H. 2003, ApJ, 599, 1238
\bibitem[Dexter, Agol \& Fragile(2009)]{dex09}Dexter, J., Agol, E., \&
Fragile, P. C. 2009, ApJ, 703, L142
\bibitem[Dexter \& Fragile(2011)]{dex11}Dexter, J., \& Fragile, P. C. 2011,
ApJ, 730, 36
\bibitem[Dexter \& Fragile(2013)]{dex12}Dexter, J., \& Fragile, P. C. 2013,
MNRAS, 432, 2252
\bibitem[Font, Ib\'a\~nez \& Papadopoulos(1998)]{fon98}Font, J. A., Ib\'a\~nez,
J. ${\rm M}^{\rm A}$., \& Papadopoulos, P. 1998, ApJ, 507, L67
\bibitem[Fragile(2009)]{fra09}Fragile, P. C. 2009, ApJ, 706, L246
\bibitem[Fragile \& Anninos(2005)]{fra05}Fragile, P. C., \& Anninos, P. 2005,
ApJ, 623, 347
\bibitem[Fragile \& Anninos(2007)]{fra07a}Fragile, P. C., \& Anninos, P. 2007,
ApJ, 665, 1507
\bibitem[Fragile et al.(2007)]{fra07}Fragile, P. C., Blaes, O. M., Anninos,
P., \& Salmonson, J. D. 2007, ApJ, 668, 417
\bibitem[Fragile \& Blaes(2008)]{fra08}Fragile, P. C., \& Blaes, O. M.
2008, ApJ, 687, 757
\bibitem[Gammie, McKinney, \& T\'oth(2003)]{gam03}Gammie, C. F., McKinney,
J. C., \& T\'oth, G. 2003, ApJ, 589, 444
\bibitem[Gillessen et al.(2012)]{gil12}Gillessen, S., Genzel, R., Fritz,
T. K., et al. 2012, Nature, 481, 51
\bibitem[Greene, Bailyn \& Orosz(2001)]{gre01}Greene, J., Bailyn, C. D., \&
Orosz, J. A. 2001, ApJ, 554, 1290
\bibitem[Henisey et al.(2009)]{hen09}Henisey, K. B., Blaes, O. M., Fragile,
P. C., \& Ferreira, B. T. 2009, ApJ, 706, 705
\bibitem[Henisey, Blaes \& Fragile(2012)]{hen12}Henisey, K. B., Blaes, O. M.,
\& Fragile, P. C. 2012, ApJ, 761, 18, 14 pp.
\bibitem[Hjellming \& Rupen(1995)]{hje95}Hjellming, R. M., \& Rupen, M. P.
1995, Nature, 375, 464
\bibitem[Ivanov \& Illarionov(1997)]{iva97}Ivanov, P. B., \& Illarionov,
A. F. 1997, MNRAS, 285, 394
\bibitem[King \& Pringle(2006)]{kin06}King, A. R., \& Pringle, J. E. 2006,
MNRAS, 373, L90
\bibitem[Krolik, Hawley, \& Hirose(2005)]{kro05}Krolik, J. H., Hawley, J. F.,
\& Hirose, S. 2005, ApJ, 622, 1008
\bibitem[Martin, Tout, \& Pringle(2008)]{mar08}Martin,
R. G., Tout, C. A., \& Pringle, J. E. 2008, MNRAS, 387, 188
\bibitem[McKinney, Tchekhovskoy, \& Blandford (2013)]{mck13}McKinney, J. C.,
Tchekhovskoy, A., \& Blandford, R. D. 2013, Science, 339, 49
\bibitem[Noble, Krolik, \& Hawley(2009)]{nob09}Noble, S. C., Krolik, J. H.,
\& Hawley, J. F. 2009, ApJ, 692, 411
\bibitem[Papaloizou \& Lin(1995)]{pap95}Papaloizou, J. C. B., \& Lin, D.
N. C. 1995, ApJ, 438, 841
\bibitem[Penna et al.(2010)]{pen10}Penna, R. F., McKinney, J. C., Narayan,
R., Tchekhovskoy, A., Shafee, R., \& McClintock, J. E. 2010, MNRAS, 408, 752
\bibitem[Sorathia, Krolik \& Hawley(2013a)]{sor13a}Sorathia, K. A., Krolik,
J. H., \& Hawley, J. F. 2013a, ApJ, 768, 133, 14 pp.
\bibitem[Sorathia, Krolik \& Hawley(2013b)]{sor13b}Sorathia, K. A., Krolik,
J. H., \& Hawley, J. F. 2013b, ApJ, in press
\bibitem[Steiner \& McClintock(2012)]{ste12}Steiner, J. F., McClintock, J. E.
2012, ApJ, 745, 136
\bibitem[Volonteri, Sikora \& Lasota(2007)]{vol07}Volonteri, M., Sikora, M.,
\& Lasota, J.-P. 2007, ApJ, 667, 704
\end{thebibliography}
\end{document}